\documentclass[aps,pre,amsmath,amssymb,twocolumn,floatfix]{revtex4}

\newif\ifpdf\ifx\pdfoutput\undefined\pdffalse\else\pdfoutput=1\pdftrue\fi

\usepackage{graphicx}
\usepackage{bm}
\usepackage{epsfig}

\newcommand{\fex}{f^{\rm ex}} 
\newcommand{\fideal}{f^{\rm id}} 
 
\newcommand{\fmom}{f_{\rm mom}}
\newcommand{\be}{\begin{equation}}
\newcommand{\ee}{\end{equation}}

\newcommand{\rh}{\rho}  
\newcommand{\sig}{\sigma} 
\newcommand{\pol}{\delta} 

\newcommand{\muex}{\mu^{\rm ex}} 

\newcommand{\mi}{\rh_i} 
\newcommand{\mze}{\rh_0}

\newcommand{\fexhs}{\fex_{\rm hs}} 

\newcommand{\rat}{\xi} 
\newcommand{\ratc}{\xi_{\rm c}} 
\newcommand{\sigp}{\sig_{\rm p}} 
\newcommand{\sigc}{\sig_{\rm c}} 

\newcommand{\mup}{\mu_{\rm p}} 

\newcommand{\rhpsig}{\rh_{\rm p}(\sigp)} 
\newcommand{\rhcsig}{\rh_{\rm c}(\sigc)} 

\newcommand{\rhp}{\rh_{\rm p}} 
\newcommand{\rhc}{\rh_{\rm c}} 

\newcommand{\wic}{w_i^{\rm c}(\sig_{\rm c})} 
\newcommand{\wip}{w_i^{\rm p}} 

\newcommand{\phic}{\phi_{\rm c}}
\newcommand{\phip}{\phi_{\rm p}}

\newcommand{\hs}{_{\rm hs}}

\begin{document}

\title{\bf Effects of colloid polydispersity on the phase behaviour of
colloid-polymer mixtures}

\author{Moreno Fasolo}
\author{Peter Sollich}
\affiliation{Department of Mathematics, King's College London, Strand, London WC2R 2LS, UK}

\date{\today}

\begin{abstract} 
We study theoretically the equilibrium phase behaviour of a mixture of
{\em polydisperse} hard-sphere colloids and monodisperse polymers,
modelled using the Asakura-Oosawa model within the free volume
approximation of Lekkerkerker {\em et al}.
We compute full phase diagrams in the plane of colloid and polymer
volume fractions, using the moment free energy method. The intricate
features of phase separation in pure polydisperse colloids combine
with the appearance of polymer-induced gas-liquid coexistence to give
a rich variety of phase diagram topologies as the polymer-colloid size
ratio $\rat$ and the colloid polydispersity $\pol$ are varied.
Quantitatively, we find that polydispersity disfavours fluid-solid
against gas-liquid separation, causing a substantial {\em lowering} of
the threshold value $\ratc$ above which stable two-phase gas-liquid
coexistence appears. Phase splits involving two or more solids can
occur already at low colloid concentration, where they may be
kinetically accessible. We also analyse the strength of colloidal size
fractionation. When a solid phase separates from a fluid, its
polydispersity is reduced most strongly if the phase separation takes
place at low colloid concentration and high polymer concentration, in
agreement with experimental observations. For fractionation in
gas-liquid coexistence we likewise find good agreement with
experiment, as well as with perturbative theories for
near-monodisperse systems.


\end{abstract} 
\maketitle


\section{\label{sec:4hspolIntroduction}Introduction}

Since the 1980s phase separation phenomena in colloidal suspensions
that are induced by the addition of polymers have gradually gained interest
both from a technological and scientific point of view.
The equilibrium phase behaviour of a pure colloidal hard sphere system
is dictated exclusively by entropic effects, and these can cause
separation into fluid and solid phases. The term fluid is used to
emphasise that there are no separate gas and liquid phases in such a
system. However, such phases can appear with the addition of polymer,
because of the effective colloid-colloid attraction which this
induces. Our aim in this paper is to study the effect in which colloid
{\em polydispersity}, i.e.\ a continuous spread of colloid sizes,
further affects the phase behaviour. Because polydispersity causes
complex phase equilibria even for pure hard
spheres~\cite{FasSol03,FasSol04}, with for example the coexistence of
a fluid with several solids, one expects very rich phase behaviour
when polymer is added.

A theoretical model of phase separation in a mixture of colloidal and
polymer particles was first advanced by Asakura and Oosawa
(AO)~\cite{AsaOos54} and extended by Vrij~\cite{Vrij76}. In order to
describe the interactions in a solution with {\it non-adsorbing}
polymers, they proposed to model each polymer chain as a sphere with
diameter $\sigp$ equal to twice the radius of
gyration. These ``polymer spheres'' are then
assumed to be able to interpenetrate freely with each other, which is
a reasonable assumption for polymers near their $\theta$-point where
the chains obey random walk statistics. On the other hand, the
polymer chains are completely excluded from the space occupied by the
colloidal particles, corresponding to a hard interaction between
polymers and colloids.

Overall, the AO model is effectively a mixture of hard spheres, but
with extreme non-additivity in the polymer-polymer interaction which
has range zero rather than $\sigp$. Explicitly, the AO interparticle
potentials can be expressed as follows:
\begin{eqnarray*}
v_{\rm cc}(r) &=& \left \{  
  \begin{array}{cl}
          \infty & {\rm if} \;  r < \frac{1}{2}({\sigc}+{\sigc'}) \\
          0 & {\rm otherwise}
  \end{array} \right. ; \\
v_{\rm cp}(r) &=& \left \{  
  \begin{array}{cl}
          \infty & {\rm if} \;  r < \frac{1}{2}({\sigc}+{\sigp}) \\
          0 & {\rm otherwise}
  \end{array} \right. ; 
\ \ v_{\rm pp}(r) = 0
\end{eqnarray*}
Here $r$ is the distance between the centres of mass of the
particles considered. The subscript ``cc'' refers to an interaction
between two colloidal particles; in a polydisperse system these can
have different diameters $\sigc$ and $\sigc'$. Similarly ``cp''
and ``pp'' denote colloid-polymer and polymer-polymer interactions,
respectively.

Even though all the interactions are still hard within this model, the
entropy of the polymers induces an effective attraction between the
colloids. 
The centres of mass of the polymers are excluded from a spherical
``exclusion zone'' of width $\sigp/2$ around the colloids. When the
exclusion zones of two or more colloids overlap, the polymers cannot
access the region between the colloids. The osmotic pressure of the
polymers is then 
unbalanced and creates an attractive {\em depletion interaction}
between the colloids. Its range is set by the polymer diameter
$\sigp$, and its strength at contact is found to be proportional to
$\sigp^2\sigc$ times the osmotic pressure of the polymers. It is this
depletion interaction which can cause the suspension to separate into
colloid-poor and colloid-rich phases, resulting in phase equilibria
with solid, liquid or gas phases depending on conditions.
These striking effects have been extensively investigated
experimentally~\cite{IleOrrPooPus95,AndDeHLek02,Poon02,TuiRieDek03,FaiEva04}
and the AO model has been shown to give a good description for
suspensions of sterically stabilised colloidal particles immersed in a
solvent with polymers~\cite{LekPooPusStrWar92,DijBraEva99}.
However, experimental systems are always polydisperse, containing a
spread of colloid sizes as well as polymer chain lengths. As reviewed
below, our theoretical understanding of the effects of this
polydispersity on the phase behaviour remains very limited. Ideally,
one would therefore like to be able to study theoretically the phase
behaviour of a mixture of colloids and polymers that are both
polydisperse.  This is a very challenging problem. As a first step we
therefore focus in this paper on the case where {\it only} the
colloids are treated as polydisperse. This already turns out to result
in very rich phase behaviour.

We first review some of the previous work carried out on
colloid-polymer mixtures. Most theoretical studies have focused on the
case where the colloids are monodisperse and the polymers are either
monodisperse or polydisperse.
\begin{figure}
  \begin{center}
  \includegraphics[width=8cm]{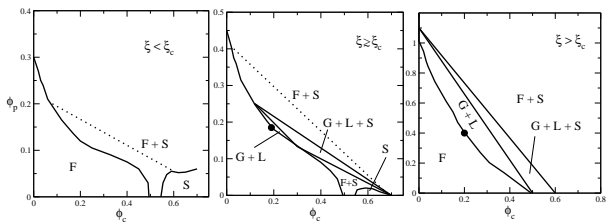}
  \caption{Phase diagram sketch for colloid-polymer mixtures with
  three different values of the size ratio $\rat$. For $\rat < \ratc$
  the depletion interaction is weak and the only effect is the
  widening of the (F)luid-(S)olid coexistence region; at $\rat>\ratc$
  (right) the longer polymers create a sufficiently long-range
  interaction that induces the formation of (G)as and (L)iquid phases
  and hence of a critical point, marked with a circle. The crossover
  value $\rat\approx\ratc$ (middle) is the value where a region of
  gas-liquid coexistence first appears.\label{fig:pha_dia_sketch}}
\end{center}
\end{figure}
%
Theoretical work by Gast {\it et al}~\cite{GasRusHal86}, Lekkerkerker
{\it et al}~\cite{LekPooPusStrWar92} and Dijkstra {\it et
al}~\cite{DijBraEva99} for a monodisperse mixture predicts that
the topology of the phase diagram depends crucially on the range
$\sigp$ of the depletion interaction, or more precisely the ratio
$\rat=\sigp/\sigc$ of the polymer and colloid diameters. This is
sketched in Fig.~\ref{fig:pha_dia_sketch}, which shows phase diagrams
in terms of colloid volume fraction $\phic$ and effective polymer
volume fraction $\phip$. The latter is defined as the polymer
concentration $\rhp$ times the volume $(\pi/6)\sigp^3$ of each
polymer and can be greater than 1 because the polymers can overlap.

To understand these phase diagrams, one can imagine gradually adding
polymers to a pure (monodisperse) colloidal hard-sphere system
which exhibits coexistence between a fluid and a crystalline solid
with volume fractions of around 50\% and 55\%,
respectively~\cite{PusVan86,PauAck90,Goetze91}.
For size ratios $\rat$ below a certain crossover value
$\ratc$ the addition of polymer only has the effect of
widening the fluid-solid coexistence region; see 
Fig.~\ref{fig:pha_dia_sketch} (left). This widening is gradual at
first but then becomes very pronounced. It has been argued that, when
viewed in terms of the colloid system with its effective
polymer-induced attraction, this can be understood as the crossover from 
a regime with dominant entropic effects to an ``energetic fluid''
regime where the strength at contact of the attractive interaction is
the key parameter~\cite{Louis01}.

For sufficiently large polymers, on the other hand ($\rat>\ratc$), the
addition of polymeric particles leads to a new phase diagram topology
where gas-liquid (and therefore gas-liquid-solid) coexistence can
occur, as shown in Fig.~\ref{fig:pha_dia_sketch} (right). The
threshold value $\ratc$ of the size ratio is the one where a
gas-liquid region first appears; see Fig.~\ref{fig:pha_dia_sketch}
(middle). The intuitive explanation for the different behaviour at
small and large $\rat$ is that if the interaction is generated by
sufficiently long polymers, $\rat>\ratc$, then neighbouring colloids
can still be within the range of the attraction even when they are in
a disordered state, and so a thermodynamically stable liquid phase can
form.

The actual crossover value $\ratc$ varies slightly depending on the
theory considered but is found to lie in the range $\ratc\approx0.3\dots0.6$.
E.g.\ in~\cite{DijBraEva99} a colloid system with the effective
AO interaction was simulated with the
result $ 0.4<\ratc<0.6$; this result is approximate because in this
range of $\rat$ the effective interaction potential for the colloids
also contains many-body terms.
Lekkerkerker {\it et al}~\cite{LekPooPusStrWar92}, by the use of a van
der Waals-type approximation (see below) identified $\ratc \approx 0.32$, and Gast
and collaborators~\cite{GasRusHal86}, with the aid of perturbation
theory, estimated $\ratc$ to be $\approx1/3$. On the experimental side
Ilett {\it et al}~\cite{IleOrrPooPus95} found instead $\ratc\approx
0.25$ for a suspension with polydispersity $\pol$ of order $0.05$.
One of the goals of our study will be to identify the effects of
colloidal polydispersity on the threshold value $\ratc$.

Initial attempts to understand the influence of polydispersity on
colloidal-polymer mixtures have been made in recent years, but with a
focus on size-polydispersity in the {\it
polymers}~\cite{Warren97,SeaFre97,LeeRob99,DenSch02} which causes
quantitative rather than qualitative changes in the phase diagrams.
To the best of our knowledge the only studies considering polydisperse
{\it colloids} investigate the liquid-gas phase separation in the
limit of near-monodisperse colloids, where perturbative theories can
be applied~\cite{EvaFaiPoo98,Evans01,FaiEva04}. In this regime it
turns out that the gas-liquid coexistence region widens as the
colloidal spheres becomes more polydisperse~\cite{Evans01}. This
statement applies when the osmotic pressure of the polymer -- or
equivalently its chemical potential -- is imposed, e.g.\ by connecting
the colloid-polymer mixture to a large polymer reservoir. This
corresponds to fixing the effective interaction between the
colloids. More recent experimental work~\cite{FaiEva04} has focused on
the strength of size fractionation between coexisting phases and its
scaling with the overall colloid polydispersity; we return to a
comparison with this study in
Sec.~\ref{sec:hspol_Fractionation_effects}.

We provide here a theoretical analysis of phase equilibria in
colloid-polymer mixtures with significant size-polydispersity in the
colloids. The paper is structured as follows. In
Sec.~\ref{sec:4hspolFreeenergychoice} we derive the free energy of the
polydisperse AO model, within the free volume approximation employed
by Lekkerkerker {\it et al}~\cite{LekPooPusStrWar92}. This free energy
has a truncatable~\cite{SolWarCat01} structure, allowing us to solve
the resulting phase equilibrium equations by using the moment free energy
(MFE) 
method as explained in
Sec.~\ref{sec:hspol_Moment_free_energy_and_implementation}. In
Sec.~\ref{sec:4hspolPhasebehaviour} we study the phase diagram
topologies that arise, while Sec.~\ref{sec:quantitative} is devoted to
a quantitative analysis of the effects of polydispersity and polymer
size on the phase diagrams, and of colloidal size fractionation
between coexisting phases. Conclusions and an outlook towards future
work are given in Sec.~\ref{sec:4hspolConclusionandoutlook}.

\section{\label{sec:4hspolFreeenergychoice}Free volume theory}

Even though in the AO model the polymer-polymer interactions are
ideal, a direct calculation of the partition function is a difficult
task because one needs to keep track of the overlaps of all exclusion
zones around the colloids; the remaining free volume accessible to the
polymers thus depends in a complicated way on the configuration of the
colloids. Even if only pairwise overlaps are considered one still
effectively has a colloid system with a depletion interaction
potential of finite range, which cannot be solved exactly. Progress
can however be made by using a van der Waals or free volume
approximation which replaces the free volume by its average over
colloid configurations. This will lead us to an explicit expression
for the Helmholtz free energy of our system as a sum of two
contributions, one corresponding to the pure hard-sphere system and
one describing the average interaction between polymers and colloids.

In what follows we will derive the free volume approximation to the
free energy for a system where both the polymers and colloids are
polydisperse. Allowing for polydispersity in the polymers adds almost
no extra work and facilitates comparison with other approaches. The
case of monodisperse polymers which we later study numerically is
easily recovered as a special case.

We assume initially that we have a discrete mixture of colloid and
polymer species and take the polydisperse limit at the end.
The colloid species are labelled by $s$, with $\{N_{{\rm c}s}\}$ the
number of particles of each species and $\sigc{}_s$ their diameter; 
similarly $\{N_{{\rm p}t}\}$ and $\sigp{}_{t}$ give the number of particles
and diameter for each of the polymer species.
Because all interactions
in the AO model
are hard, temperature becomes an irrelevant scale factor which only
sets the energy scale. We therefore measure all energies in units of
$T=1/\beta$ and set $k_B$ to one.
The spatial coordinates of the particles are denoted by $R_i$ for the colloids
and $r_j$ for the polymers. The canonical partition function is then
\begin{eqnarray}
Z &=& \prod_{s,t} \frac{\lambda_{{\rm c}s}^{-3 N_{{\rm c}s}} 
\lambda_{{\rm p}t}^{-3 N_{{\rm p}t}}}{N_{{\rm c}s}! \; N_{{\rm p}t}!} 
\int dR  \;  dr  \; 
\nonumber \\
& &\times
\exp\left[- \sum_{i<i'}v_{\rm cc}(i,i') - \sum_{i,j} v_{\rm cp}(i,j)\right]
\label{eq:z_can}
\end{eqnarray}
where $v_{\rm cc}(i,i')$ stands for $v_{\rm cc}(|R_i-R_{i'}|)$ and
similarly for $v_{\rm cp}(i,j)$. The integrations $dR$, $dr$ extend
over all colloid and polymer coordinates, and $\lambda_{{\rm c}s}$ and
$\lambda_{{\rm p}t}$ are the de Broglie wavelengths for each
species. The natural thermodynamic potential to work with, in this
context, is the semi-grand canonical potential. For this the system can
be thought of as connected to a polymer reservoir via a membrane
impermeable to colloids but permeable to polymers,
so that the polymer particle numbers $\{N_{{\rm p}t}\}$ can range over
all possible values. The resulting semi-grand partition function is
given by
\begin{equation}
\Xi(\{N_{{\rm c}s}\},V,\{\mup{}_t\}) = 
\sum_{\{ N_{{\rm p}t}\}} \prod_t e^{\mup{}_t N_{{\rm p}t}}
Z(\{N_{{\rm c}s}\},\{N_{{\rm p}t}\},V)
\label{eq:z_sgcan}
\end{equation}
where the chemical potentials $\mup{}_t$ of the polymers are fixed by
the reservoir.
In (\ref{eq:z_sgcan}) we can separate the ideal contributions from
the excess part:
\begin{eqnarray}
\Xi &=&  Z^0\hs \sum_{ \{ N_{{\rm p}t} \}} 
\int \frac{dR}{V^{N_{\rm c}}} \;  
\exp\left(- \sum_{i<i'}v_{\rm cc}(i,i')\right)
\nonumber
\\
& & \times\prod_t \frac{V^{N_{{\rm p}t}}}{N_{{\rm p}t}!}\left(\int
\frac{dr_{j(t)}}{V} 
\frac{e^{ \mup{}_t}}{\lambda_{{\rm p}t}^3}
e^{-  \sum_{i,j} v_{\rm cp}(i,j)} \right)^{N_{{\rm p}t}} 
\label{eq:par_can}
\end{eqnarray}
where $Z^0\hs$ is the canonical partition function of an ideal mixture
of colloids, $Z^0\hs=V^{N_{\rm c}}\prod_{\rm s}{\lambda_{{\rm c}s}^{-3
N_{{\rm c}s}}}/{N_{{\rm c}s}!}$ and $N_{\rm c}$ is the total number of
colloids. 
The quantity ${e^{\mup{}_t}}/{\lambda_{{\rm p}t}^3}$ identifies the
thermodynamic {\it activity} of the polymer species $t$, which for our
ideal polymers is just its density $\rho^{\rm r}_{{\rm p}t}$ in the
reservoir.
The integration over $r_{j(t)}$, which represents the position of any
one of the polymers from species $t$, gives the fraction of volume
available to a hard sphere of diameter $\sigp{}_t$ if the colloids are
in a configuration $\{R_i\}$,
$\alpha(\{R_i\},\sigp{}_t)$ $=$ $V^{-1}\int dr_{j(\rm t)} e^{-
\sum_{i} v_{\rm cp}(|R_i-r_j|)}$. We are thus left with a
configurational integral over the hard spheres:
\begin{eqnarray}
\Xi &=& Z^0\hs
\int \frac{dR}{V^{N_{\rm c}}} \exp{\left(-\sum_{i<i'}v_{\rm
cc}(i,i')\right)} \nonumber\\ 
& &\times \exp\left({V \sum_t \rho^{\rm r}_{{\rm p}t} \; \alpha(\{R_i\},
\sigp{}_t)}\right) \; .
\label{eq:con_int_pol}
\end{eqnarray}
Following Widom~\cite{Widom63} this can be reinterpreted as an average
$\langle\;\rangle\hs$ over the Boltzmann distribution of the pure
colloidal hard-sphere system, with excess partition function $Z^{\rm
ex}\hs$:
\begin{equation}
\Xi = Z^0\hs Z^{\rm ex}\hs
\left\langle \exp\left(V \sum_t \rho^{\rm r}_{{\rm p}t} \; \alpha(\{R_i\},
\sigp{}_t)\right) \right\rangle\hs 
\label{eq:hspol_can_ave}
\end{equation}
%
We note that the second exponential in (\ref{eq:con_int_pol}) is the
effective colloid-colloid interaction generated by the
polymers. Beyond trivial zero- and one-body terms it always contains a
two-body contribution which is the attractive depletion interaction
described above. For sufficiently large polymers ($\rat>0.154$) also
interactions between three or more colloids appear because more than
two exclusion zones can overlap simultaneously.

The average in~(\ref{eq:hspol_can_ave}) cannot be evaluated
exactly. To make progress, we follow the van der Waals approach or
``free volume theory'' of Lekkerkerker {\it et
al}~\cite{LekPooPusStrWar92}. This consists of ``moving'' the average
$\langle\;\rangle\hs$ inside the exponential to give
\begin{equation}
\Xi = Z^0\hs Z^{\rm ex}\hs
\exp\left(V \sum_t \rho^{\rm r}_{{\rm p}t} \; \left\langle \alpha(\{R_i\},
\sigp{}_t)\right\rangle\hs\right) \;. 
\end{equation}
This approximation is exact in the limit of small polymer activities
$\rho^{\rm r}_{{\rm p}t}\to0$; otherwise it can be viewed as identifying a
lower bound on $\Xi$, or an upper bound on the thermodynamic potential
$\Omega=-\ln\Xi$. The latter becomes
\begin{equation}
\Omega = 
F\hs - 
V \sum_t \rho^{\rm r}_{{\rm p}t} \, \left\langle \alpha(\{R_i\},
\sigp{}_t)\right\rangle\hs
\end{equation}
where $F\hs$ is the Helmholtz free energy of the pure colloid
system. The average free volume $\left\langle \alpha(\{R_i\},
\sigp{}_t)\right\rangle\hs$ which appears here is the probability
of being able to insert a hard particle of diameter $\sigp{}_t$ into
the pure colloid system. But from the Widom insertion
principle~\cite{Widom63} 
this is just $\exp[-\muex\hs(\sigp{}_t)]$ where $\muex\hs(\sig)$ is the excess chemical potential of a {\em colloid} particle
of diameter $\sig$, in a system without polymer. The semi-grand potential 
per unit volume, $\omega=\Omega/V$, can thus be written as
%
\begin{equation}
\omega = 
f\hs -
\sum_t \rho^{\rm r}_{{\rm p}t} \;  e^{-\muex\hs(\sigp{}_t)}
\label{eq:omega_discrete}
\end{equation}
%
In the polydisperse limit, the $\rho^{\rm r}_{{\rm p}t}$ turn into a
reservoir polymer density distribution $\rhp^{\rm r}(\sigp)$ and the
sum over $t$ into an integral over $\sigp$
so that
\begin{equation}
\omega = 
f\hs - 
\int d\sigp\,  \rhp^{\rm r}(\sigp)  e^{-\muex\hs(\sigp)}
\label{eq:omega_semigrand}
\end{equation}
%
We note that a similar expression was derived in~\cite{DenSch02} for
the case of monodisperse colloids. This approach used fundamental
measure theory and leads to the scaled particle-theory expressions for
$f\hs$ and $\muex\hs(\sig)$. Only colloidal gas-liquid
demixing was investigated, and the polymer polydispersity was
interpreted as arising from the compressibility of polymer chains of a
given length. A semi-grand approach, with the slight modification of
keeping the total polymer number fixed, is then appropriate because
the different polymer ``species'' can effectively be transformed into
each other.

For genuine polymer chain length polydispersity, however, we are in
practice dealing with a sample containing a fixed number of polymers
of each species. We therefore need to transform back to the canonical
ensemble to obtain the Helmholtz free energy density, $f=\omega+\sum_t
\mup{}_t \rhp{}_t$.  From~(\ref{eq:omega_discrete}) and bearing in
mind the definition of $\rho^{\rm r}_{{\rm p}t}$, the polymer
densities and chemical potentials are related by the following
transformation:
\begin{equation}
\rhp{}_t = -
\frac{\partial\omega}{\partial\mup{}_t} = 
\lambda_{{\rm p}t}^{-3} e^{\mup{}_t}
e^{-\muex_{\rm c}(\sigp{}_t)} =
\rho^{\rm r}_{{\rm p}t} e^{-\muex_{\rm c}(\sigp{}_t)} 
\end{equation}
Simple algebra then gives for the free energy density
\begin{equation}
f = f\hs + \sum_t \rhp{}_t(\ln\rhp{}_t-1)
+ \sum_t \rhp{}_t \muex_{\rm c}(\sigp{}_t)
\label{eq:f_discrete}
\end{equation}
Here we have separated off the ideal polymer part in the second term,
dropping the contribution $\sum_t \rhp{}_t \ln
\lambda_{{\rm p}t}^3$ which is linear in the densities and therefore
leaves the phase behaviour unaffected. In the limit of fully
polydisperse polymers the free energy (density) would then read
\begin{eqnarray}
f & = &
f\hs + 
\int d\sigp\, \rhpsig \left[\ln \rhpsig - 1\right]\nonumber \\
& &{}+{}\int d\sigp\, \rhpsig \muex\hs(\sigp)
\label{eq:pol_fre_ene}
\end{eqnarray}
%
However, in the present study we neglect polydispersity of the
polymers and focus on the effects of colloid polydispersity. After
separating the free energy $f\hs$ of the pure colloid system into
its ideal and excess parts, the total free
energy~(\ref{eq:f_discrete}) then simplifies to $f = \fideal + \fex$,
with
\begin{eqnarray}
\fideal &=&
\int d\sigc\, \rhcsig [\ln \rhcsig-1] +  \rhp (\ln\rhp-1)\nonumber\\
\fex &=& \fexhs + \rhp{} \muex\hs(\sigp)\; .
\label{eq:f_monopol}
\end{eqnarray}

Importantly, evaluation of the free energy~(\ref{eq:f_monopol})
requires as nontrivial input only the properties of the pure colloid
system, i.e.\ its excess free energy $\fexhs$ and the excess chemical
potentials $\muex\hs(\sig)$ which are obtained by
differentiation, $\muex\hs(\sig)=\delta \fexhs/\delta\rhc(\sig)$. To
specify the free energy fully, we therefore only need to assign
appropriate expressions for $\fexhs$ in the colloidal fluid (or
gas/liquid) and solid phases. For the {\it fluid} part of the excess
free energy the most accurate approximation available is the
generalisation by Salacuse and Stell~\cite{SalSte82} of the BMCSL
equation of state~\cite{Boublik70,ManCarStaLel71} while for the {\it
solid} we adopt Bartlett's fit to simulation data for bidisperse hard
sphere mixtures~\cite{Bartlett99,Bartlett97}. Our previous 
work~\cite{FasSol03,FasSol04}
on polydisperse hard spheres has shown that with these free energy
expressions quantitatively accurate fits to simulation data are
obtained, and so we continue to use them for the present study.

Some care is needed in the application of the excess free energy
expression for the solid. The simulation data from which this is
derived were obtained for mixtures of two species of hard spheres with
diameters differing by no more than $\approx 15\%$. The resulting
excess chemical potentials $\muex_{\rm c}(\sig)$ are therefore
accurate only within a small range around the mean colloid
diameter. Outside this range they are unreliable; for example the
limiting behaviour predicted from the Widom insertion principle for
$\sig\to 0$ is not retrieved correctly. This causes a difficulty
because in the colloid-polymer interaction term in the free
energy~(\ref{eq:f_monopol}) we need the excess chemical potential
evaluated at the {\em polymer} diameter, which we will take to be
rather smaller than the mean colloid diameter. In fact, it is only in
this regime of size ratios $\rat$ well below unity that the AO-model
approximation of treating polymers as spherical particles is
reasonable; for larger polymers the colloids can ``see'' the polymer
chain structure~\cite{MonLouBolRot03}.

To circumvent this problem, we follow previous work within the free
volume approach~\cite{LekPooPusStrWar92,Warren97,SeaFre97} and always
evaluate the excess chemical potential $\muex_{\rm c}(\sigp)$
governing the polymer-colloid interaction from the BMCSL free
energy. The underlying physical approximation is that the free volume
available to the polymers is not drastically different for colloidal
fluid and solid phases so that the {\em fluid} (BMCSL) expression can
also be used to estimate the free volume in colloidal {\em solids}.

\section{Moment free energy method
\label{sec:hspol_Moment_free_energy_and_implementation}
}

A key feature of the excess free energies which we use to describe the
colloidal fluid and solid phases is that they are {\em
truncatable}~\cite{SolWarCat01}: they depend only on the finite set of
moments $\mi=\int d\sigc\rhcsig \sigc^i$ ($i=0,\ldots, 3$) of the
colloidal density distribution $\rhcsig$. Here $\rhcsig d\sigc$ is the
number density of colloids with diameters in the range
$\sigc\ldots\sigc+d\sigc$. The excess chemical potentials are then
third order polynomials,
\begin{equation}
\muex\hs(\sig) = \frac{\delta \fex\hs}{\delta\rhc(\sig)} =
\sum_{i=0}^3 \muex_{{\rm hs,}i} \sig^i
\end{equation}
The coefficients $\muex_{{\rm hs},i}$ here are the moment excess
chemical potentials of a pure hard sphere system, $\muex_{{\rm hs},i} =
\partial
\fex\hs/\partial \mi$, and also depend only on the $\mi$. The
excess part of the free energy~(\ref{eq:f_monopol}) of our interacting
colloid-polymer mixture thus takes the form
\begin{equation}
\fex = 
\fex\hs(\{\mi\})  + \rhp  \sum_i \muex_{{\rm hs,}i}
\sigp^i \; .
\label{eq:fre_ene_mix}
\end{equation}
Recall that in the last term of~(\ref{eq:fre_ene_mix}), i.e.\ in the
polymer-colloid interaction, we always use the $\muex_{{\rm hs,}i}$
derived from the BMCSL free energy.

The excess free energy~(\ref{eq:fre_ene_mix}) depends only on a finite
number of variables, namely the $\mi$ and the polymer density $\rhp$.
Importantly, these can again be viewed as moments of an
{\em enlarged} density distribution which collects all the densities of our
system, namely $(\rhcsig,\rhp)$. Specifically, we define moments by
\begin{equation}
\mi=\int d\sigc\,\rhcsig\wic+\rhp\wip
\label{eq:enlarged_moment_def}
\end{equation}
in terms of weights $(\wic,\wip)$ which are made up of a weight
function $\wic$ for the colloid part and a single coefficient $\wip$
for the polymer part. The colloidal moment densities, $\mi$,
($i=0,\ldots,3$) are then given by the weight functions $(\sigc^i,0)$
while the polymer density $\rhp\equiv \rho_4$ is the moment with
weight $(0,1)$.
In summary, our system is described by an excess free energy which
depends on five moment densities of the enlarged density distribution
$(\rhcsig,\rhp)$ and is therefore by definition truncatable. This
allows us to employ the moment free energy method for the calculation
of phase diagrams. As described
in~\cite{SolWarCat01,Warren98,SolCat98,Sollich02}, and for the
specific case of polydisperse hard spheres in~\cite{FasSol04},
this maps the full free energy~(\ref{eq:f_monopol}), with its
dependence on all details of $(\rhcsig,\rhp)$ through the ideal part,
onto a moment free energy (MFE) $\fmom(\{\rho_i\}) = \left(\sum_i
\lambda_i\mi - \rho_{\rm tot}\right)+\fex(\{\mi\})$
which depends only on the moments $\mi$ ($i=0,\ldots,4$). Here
$\rho_{\rm tot}=\mze+\rho_4$ is the total number density of colloids
and polymers and the $\lambda_i$ are Lagrange multipliers which depend
implicitly on the values of the $\{\mi\}$.  From this MFE, phase
behaviour can then be found by the conventional methods for finite
mixtures, treating each of the $\mi$ as a number density of an
appropriate quasi-species. For truncatable free energies this
locates exactly the cloud points, i.e.\ the onset of phase separation,
as well as the properties of the coexisting ``shadow'' phases that
appear there. Inside the coexistence region, one in principle needs to
solve a set of highly coupled nonlinear equations and the predictions
derived from the MFE are only approximate. However, by retaining extra
moments with adaptively chosen weight functions, increasingly accurate
solutions can be obtained by
iteration~\cite{ClaCueSeaSolSpe00,SolWarCat01}. Using these as initial
points, we are then able to find full solutions of the exact -- for
our model free energy -- phase equilibrium
equations~\cite{SpeSol02,SpeSol03a}.

We work with dimensionless units in the following. We call the overall
colloid density distribution in the system the ``parent'' colloid
distribution, $\rho_{\rm c}^{(0)}(\sigc)$, with overall colloid
density $\rho_{\rm c}^{(0)}= \int d\sigc\,\rho_{\rm
c}^{(0)}(\sigc)$. The normalised parent size distribution is $n_{\rm
c}^{(0)}(\sigc)= \rho_{\rm c}^{(0)}(\sigc)/\rho_{\rm c}^{(0)}$; we
denote its mean diameter by $\sig_0$. Lengths are measured in units of
$\sig_0$; the dimensionless polymer diameter $\sigp$ then coincides
with the polymer-to-(mean) colloid size ratio $\rat$. All densities are
referred to the volume of a unit colloid particle, $(\pi/6)\sig_0^3$.

\section{\label{sec:4hspolPhasebehaviour}Phase diagram topologies}

In this section we will describe our results for the overall phase
behaviour of a mixture of polydisperse hard spheres and monodisperse
polymers. Our numerical work requires a choice to be made for the
colloidal parental diameter distribution. We focus on
a triangular distribution,
\[
n_{\rm c}^{(0)}(\sigc) = \frac{1}{w^2} \left\{
\begin{array}{llrcl}
\sig-(1-w) \quad & \mbox{for} & 1-w  \leq \sigc \leq  1 \\
(1+w)-\sig \quad & \mbox{for} & 1  \leq \sigc \leq  1+w
\end{array}\right.
\]
whose width parameter $w$ is related to the polydispersity by
$w=\sqrt{6}\,\pol$. For the moderate values of $\pol$ of interest here
one expects other distribution shapes to give qualitatively similar
results. As in the case of polydisperse hard spheres without added
polymer, this is based on the intuition that for narrow size
distributions $\pol$ is the key parameter controlling the phase
behaviour~\cite{Pusey87,FasSol03,FasSol04}.

To assess the range of possible phase diagram topologies we consider
initially four combinations resulting from having either small or
large polymers, $\xi=0.2$ and 0.4, and colloids with small or
moderate polydispersity, $\pol=0.05$ and 0.08.
\begin{figure}
  \begin{center}
  \includegraphics[width=8cm]{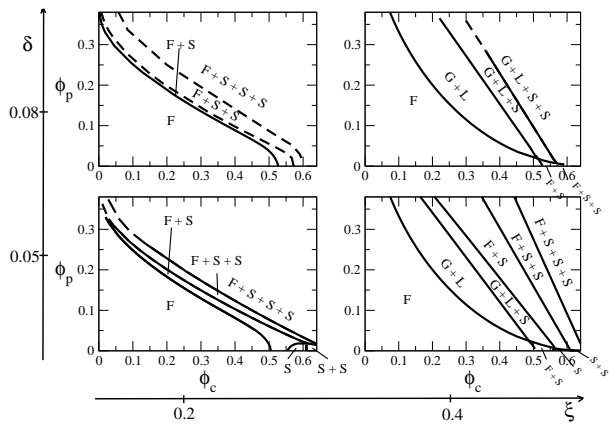}
  \caption{Phase diagram on a $2\times 2$ grid of values of
$(\xi,\pol)\in\{0.2, 
  0.4\}\times\{0.05, 0.08\}$. In each region the nature of the
  phase(s) coexisting at equilibrium is indicated (F: fluid, S: solid,
  G: gas, L: liquid). The dashed lines indicate the best guess of the
  phase boundary in regions where our numerical data become
  unreliable.
\label{fig:xi_d}}
  \end{center}
\end{figure}
Fig.~\ref{fig:xi_d} shows the resulting phase diagrams in the
$\phic$-$\phip$ plane. The nature of the different coexisting phases,
i.e.\ gas, liquid, fluid and solid, is indicated in each region.
Probably the most striking feature of these phase diagrams is the
presence of multiple solids: in addition to the phases that are found
for monodisperse colloids (see Fig.~\ref{fig:pha_dia_sketch}), size
polydispersity in the colloids allows the system to phase separate
into several solid phases. 

\begin{figure}
  \begin{center} \includegraphics[width=8cm]{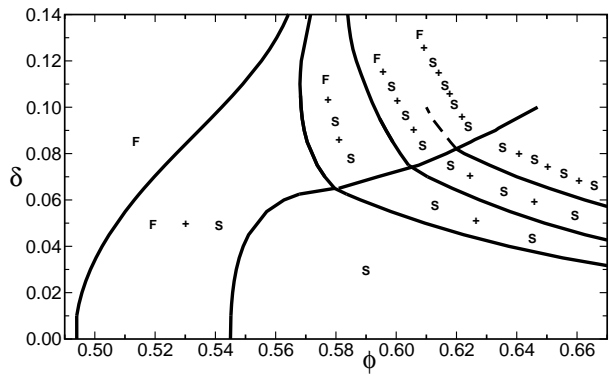}
  \caption{Phase diagram for polydisperse hard spheres without added
  polymer. Note the appearance of regions with multiple solid
  coexistence as polydispersity increases. Horizontal cuts at
  $\pol=0.05$ and $0.08$ give the behaviour along the baselines of
  the graphs in the previous
  figure. From~\protect\cite{FasSol03}.
\label{fig:parent_hs}}
  \end{center}
\end{figure}

Qualitatively, the phase diagrams of Fig.~\ref{fig:xi_d} can be
understood by combining the behaviour observed in two simpler cases:
the limit of a mixture of polymers and {\em monodisperse} colloids,
which gives the phase diagrams sketched in
Fig.~\ref{fig:pha_dia_sketch}, and the polymer-free limit of
polydisperse hard-sphere colloids~\cite{FasSol03,FasSol04}, with
phase diagram shown in Fig.~\ref{fig:parent_hs}.
%
The first limit clarifies why we need to distinguish between small and
large polymer sizes: the short- or long-range nature of the depletion
interaction dictates whether or not gas-liquid phase separation
occurs. 
The second case of pure colloids clarifies the role of polydispersity
in the phase behaviour. In fact the baseline ($\phip=0$) of each
diagram in Fig.~\ref{fig:xi_d} corresponds to a horizontal cut at the
appropriate colloid polydispersity $\pol$ in the polymer-free phase
diagram of Fig.~\ref{fig:parent_hs}. The multiple solid phases
occurring in Fig.~\ref{fig:xi_d} are thus ``inherited'' from the phase
behaviour of the polydisperse colloid system without polymer.

Having explained the broad intuition behind Fig.~\ref{fig:xi_d}, let
us look at the individual phase diagrams in more detail.
Consider first the case with small polymers, $\xi=0.2$, and moderate
colloid polydispersity, $\pol=0.05$. From Fig.~\ref{fig:parent_hs} we
see that as we move along a dilution line at $\pol=0.05$, i.e.\ as we
increase the colloid density at fixed normalised parent size distribution, we
have the following sequence of phase splits: F $\to$ F+S $\to$ S $\to$
S+S $\to \dots$. This sequence re-occurs on the baseline of the bottom left
graph of Fig.~\ref{fig:xi_d} as it must. If we now gradually add
polymers to the solution we find that, as in the monodisperse case,
the F+S coexistence region becomes wider; but in the polydisperse case,
the boundary between the F+S and S regions eventually meets that
between the S and S+S regions, resulting in a triple point that
marks the beginning of an F+S+S coexistence region. A phase split of
this type is not present in the polymer-free system at this
polydispersity $\pol$ and results from the interplay of the
polymer-generated attraction force, which favours fluid-solid phase
splits over single-phase solids, and the polydispersity of the
colloids. At higher values of polymer or colloid concentration,
coexistence of a fluid with an increasing number of solids then
occurs.

At higher colloid polydispersity, $\pol = 0.08$, the phase diagram
topology is different. On the $\phip=0$ baseline the phase split
sequence now consists of a fluid phase coexisting with an increasing
number of solid phases, at least up to $\phic\approx 0.63$. As
Fig.~\ref{fig:xi_d} (top left) shows, all these phase boundaries are
affected by the addition of polymer in the same way as the boundary
between F and F+S in the monodisperse case, shifting to smaller
$\phic$ as $\phip$ is increased. There are no phase boundaries at
which a fluid phase is lost and which would move to {\em larger} $\phic$, as
was the case for the F+S $\to$ S boundary at $\pol=0.05$. Consequently
no phase boundaries meet with increasing $\phip$ and the phase split
sequence remains as for the polymer-free system.

Using similar arguments it is possible to explain the topologies at
$\xi>\ratc$. In the system with polydispersity $\pol=0.05$ shown in
Fig.~\ref{fig:xi_d} (bottom right), we see again that the
polymer-induced attraction favours fluid-solid splits. As the polymer
concentration is increased from the baseline -- which is the same as
in Fig.~\ref{fig:xi_d} (bottom left) -- the phase equilibria therefore
acquire an additional fluid phase, via transitions from F to G+L, from
F+S to G+L+S and from S to F+S. Because of the colloid
polydispersity, however, the same mechanism now also causes
coexistence of a fluid with several solids, with e.g.\ S+S becoming
F+S+S. At higher level of polydispersities ($\pol = 0.08$,
Fig.~\ref{fig:xi_d} (top right)), all phase splits on the baseline
already contain a fluid phase which splits into G+L on adding
polymer. In particular, this mechanism turns F+S+S into G+L+S+S, a
polydispersity-induced coexistence of gas, liquid, and two solids.

\begin{figure}
  \begin{center}
  \includegraphics[width=8cm]{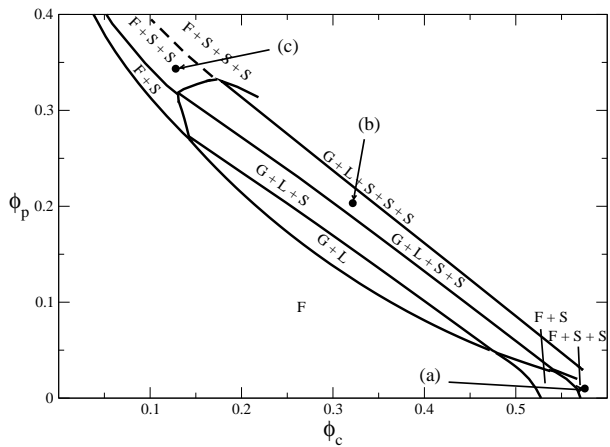}
  \caption{Phase diagram at $\xi=0.3$ and $\pol=0.08$. Note the
  topology at high polymer concentrations where the region of
  gas-liquid phase splits terminates, giving way to phase separation
  involving only a single fluid phase and one or more solids. The
  arrows indicate at which points in the phase diagram the diameter
  distributions in Fig.~\ref{fig:hspol_example_dis} below are calculated.
\label{fig:high_pol_rho}}
  \end{center}
\end{figure}

What remains unclear in this last phase diagram (Fig.~\ref{fig:xi_d} (top
right)) is how the phase boundaries meet at high polymer
concentration. In order to understand this we explored an
intermediate polymer size $\rat=0.3$ at the same polydispersity
$\pol=0.08$. This is easier computationally than exploring higher
polymer densities at $\rat=0.4$ because the smaller value of
$\rat=0.3$ moves the interesting region of the phase diagram to lower
$\phip$. As Fig.~\ref{fig:high_pol_rho} demonstrates, the region
involving gas-liquid separation eventually terminates at high density,
forming a closed ``loop''; above this loop one has a single fluid
phase coexisting with one or multiple solid phases.

\begin{figure*}
  \begin{center}
  \includegraphics[width=16cm,height=10cm]{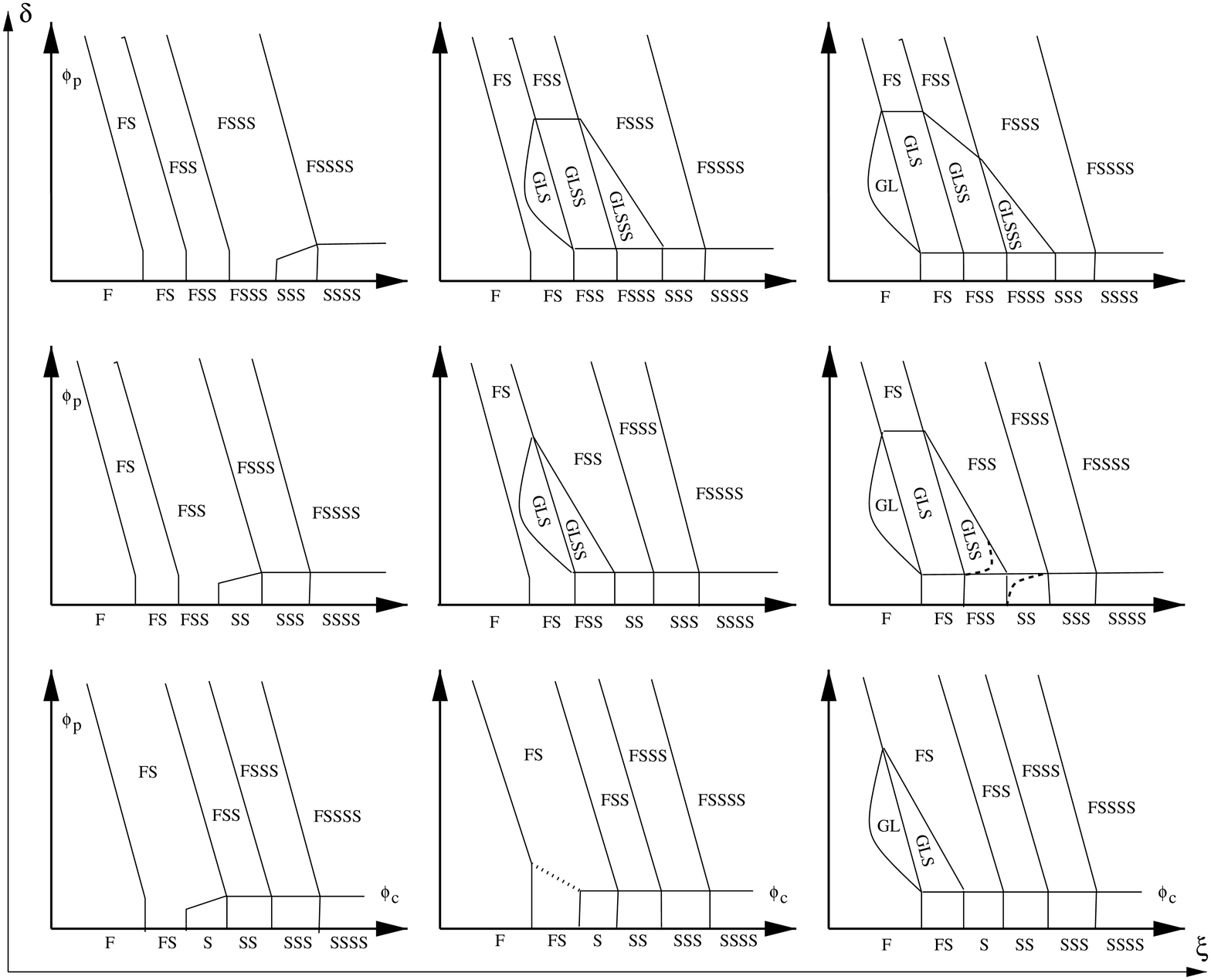}
  \caption{Sketch of expected phase diagram topologies as a function
  of polymer size $\rat$ and colloid polydispersity $\pol$. The
  sequence of phases along the $\phip=0$-baseline in each row
  corresponds to $\pol=0.05$, 0.07 and 0.08 respectively. Middle row,
  right: Dashed lines indicate a possible alternative topology which
  can still be connected smoothly to the one below but is physically
less plausible. Bottom row, middle: The dotted line indicates the
remnant of the G+L+S three-phase region, whose area shrinks to zero at
the transition between the two topologies on the left and right.
\label{fig:topologies}
}
  \end{center}
\end{figure*}
The phase diagram topologies that we have shown so far are the most
common that we have encountered, but are not the only ones possible.
This is clear from two arguments. First, we have not exhausted all
possible sequences of phase splits along the polymer-free baseline;
for colloid polydispersity $\pol=0.07$, for example,
Fig.~\ref{fig:parent_hs} shows that the sequence is F $\to$ F+S $\to$
F+S+S $\to$ S+S $\to$ S+S+S $\to$ \ldots. Second, at constant colloid
size distribution the phase diagrams need to change smoothly as the
polymer size $\rat$ is varied, connecting e.g.\ the phase diagrams on
the left of Fig.~\ref{fig:xi_d} to those on the right. In
Fig.~\ref{fig:topologies} we show a sketch of the various topologies
that should result as colloid polydispersity $\pol$ and polymer size
$\rat$ are varied. The usual rule that exactly one phase must be lost
or gained on crossing a boundary and the condition of smooth variation
with $\rat$ and $\pol$ constrain the possible topologies to a large
extent. They do not, however, make them fully unique, and
Fig.~\ref{fig:topologies} shows a possible alternative topology for
one of the phase diagrams (middle right). Nevertheless, we believe
that the scenario shown in Fig.~\ref{fig:topologies} is the most
physically plausible sequence of phase diagram topologies.  The
topology shown in the top right corresponds to
Fig.~\ref{fig:high_pol_rho} and the top right of Fig.~\ref{fig:xi_d},
although the numerically calculated phase diagrams do not extend
sufficiently far to see all the different phase splits expected
theoretically. The remaining phase diagrams in Fig.~\ref{fig:xi_d}
correspond similarly to top left and bottom left and right of
Fig.~\ref{fig:topologies}.

One intriguing feature of the predicted topologies is that phase
splits involving gas and liquid phases can persist even after the
region of pure G+L phase equilibrium has vanished from the phase
diagram. This is visible in the middle and top rows of
Fig.~\ref{fig:topologies}. As $\xi$ is decreased, the gas-liquid loop
retracts relative to the onset of fluid-solid phase separation. At the
value of $\xi$ where the two boundaries cross, the area of the
G+L-region has shrunk to zero. However, a three-phase G+L+S-region of
finite size survives at this point, and cannot disappear
discontinuously as $\xi$ is varied. Thus phase equilibria involving
gas and liquid phases can occur even if the {\em initial} phase
separation that occurs on increasing density is always into fluid and
solid phases. This effect is possible only because of the size
polydispersity of the colloids.

\section{Quantitative analysis of phase diagrams}
\label{sec:quantitative}

So far we have discussed the possible topologies of the phase diagrams
of our colloid-polymer mixture. In the present section we will study
their quantitative dependence on the colloid polydispersity as well as
on the polymer-colloid size ratio $\rat$.
We first analyse the behaviour of the cloud and shadow curves which
define the onset of phase coexistence, and consider the dependence of
the crossover value $\ratc$ of the polymer size on colloid
polydispersity. Fractionation effects are analysed next, both for
fluid-solid and for gas-liquid phase separation.  Finally we study the
shape of the internal phase boundaries between regions with two or
more phases, showing how some general features can be predicted with
simple arguments.

\subsection{Cloud and shadow curves}
\label{sec:cloud_shadow}

\begin{figure}
  \begin{center}
  \includegraphics[width=8cm]{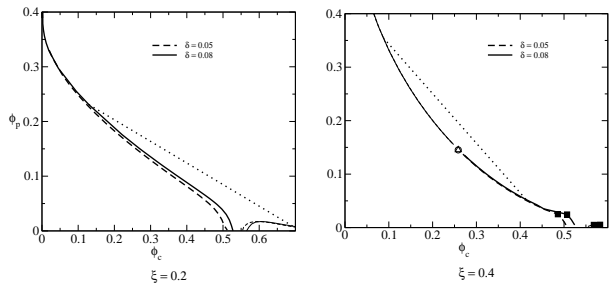}
  \caption{Plots of cloud (thick line) and shadow (thin line) curves
  at $\pol=0.05$ and $0.08$. Dotted lines connect example cloud-shadow
  pairs. Left: small polymers, $\rat=0.2$; here the cloud curve marks
  the onset of fluid-solid phase separation. Right: large polymers,
  $\rat=0.4$. The cloud curve at the larger $\phip$ gives the onset of
  gas-liquid separation. Because cloud and shadow curves at either
  side of the critical point (marked by the circle and triangle) are
  almost indistinguishable, we plot only the cloud curve below the
  critical point and the corresponding shadow above it. At low $\phip$
  one has onset of fluid-solid coexistence; the squares mark the
  triple points where the two branches of the cloud curve meet.
\label{fig:d_xi_phip_vs_phic}}
  \end{center}
\end{figure}

The cloud curve defines the onset of phase separation coming from a
single-phase region, while the shadow curve records the properties of
the incipient phase at this point. We now ask how these curves are
affected quantitatively by colloid polydispersity. Looking back at
Fig.~\ref{fig:xi_d}, it is clear that at fixed $\rat$ any such effects
on the {\em cloud curves} are small.  In
Fig.~\ref{fig:d_xi_phip_vs_phic} we show the cloud curves together
with the shadows at $\pol=0.05$ and $\pol=0.08$.
%

Also the shadow curves are seen to be only weakly affected by
polydispersity. The dotted lines connecting cloud and shadow pairs
demonstrate that, as in the monodisperse case, there is strong
colloid-polymer fractionation at the onset of phase separation, with
the polymer preferably found in the lower-density phase.
However, polydispersity does introduce an additional feature, namely
colloid size fractionation; this is discussed in the next subsection.
We consider next the quantitative effects of polydispersity and
polymer size on the onset of gas-liquid phase separation as shown in
Fig.~\ref{fig:d_xi_phip_vs_phic} (right). To do so, we calculate the
location of the gas-liquid critical point and the associated
gas-liquid spinodal. These quantities can be obtained in a
straightforward way, and without approximation, from the MFE. The
conceptual reason for this is that the local stability of a system can
be evaluated by only checking the effects of perturbations along the
moments contained in the excess part of its free
energy~\cite{SolWarCat01}.
%
\begin{figure}
  \begin{center}
\includegraphics[width=8cm]{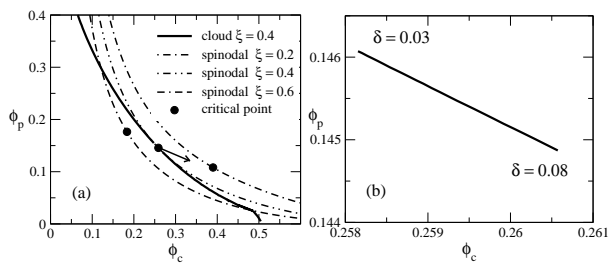}  
 \caption{(a) Spinodal curves obtained for different values of $\rat$
 together with the cloud curve at $(\rat=0.4,\pol=0.05)$ and critical
 point. (b) Change in the position of the critical point with colloid
 polydispersity $\pol$, at fixed polymer size $\rat=0.4$.
\label{fig:cp_all}}
  \end{center}
\end{figure}
The results are shown in Fig.~\ref{fig:cp_all}(a) where we plot the
previously derived cloud curve at $\rat=0.4$ and $\pol=0.05$ together
with spinodals and critical points obtained at different values of the
polymer size $\rat$. The dependence on $\rat$ is pronounced. In
Fig.~\ref{fig:cp_all}(b) we contrast this with the change in the
location of the critical point as $\pol$ rather than $\rat$ is
varied. The effect is very much smaller, and would be invisible on
the scale of Fig.~\ref{fig:cp_all}(a). We can nevertheless indicate
the {\em direction} of the shift in the critical point with increasing
$\pol$. This is seen to point inwards, which means that colloid
polydispersity tends to delay the onset of gas-liquid separation.

We note briefly that the shift in the critical point coordinates
$\phic$ and $\phip$ is quadratic in $\pol$ for the modest
polydispersities studied
here.
This arises because the critical point condition only involves moments
of the size distribution, all of which are shifted by amounts of
$O(\pol^2)$ for small $\pol$.
The dependence of the phase boundaries on $\rat$, on the other hand,
is essentially identical to that observed for monodisperse
colloids. The spinodals -- and, by inference, the cloud curves --
shift to the right as $\rat$ decreases; eventually, at the threshold
value $\rat=\ratc$, the G+L phase separation then becomes metastable
with respect to the F+S transition.

\begin{figure}
  \begin{center}
  \includegraphics[width=8cm]{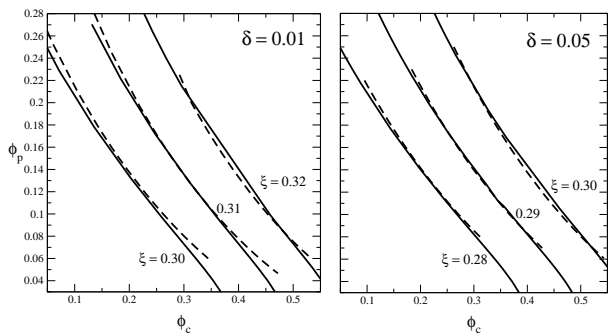}
  \caption{Plot of the phase boundaries defining the onset of F+S
  (solid line) and G+L (dashed line) coexistence, at colloid
  polydispersities $\pol=0.01$ (left) and $\pol=0.05$ (right). To make
  the curves at different polymer sizes $\rat$ visually
  distinguishable, those for the lowest and highest $\rat$-values have been
  shifted along the horizontal $\phic$-axis by $-0.1$ and $0.1$
  respectively.
\label{fig:xi_crossover}}
  \end{center}
\end{figure}
%
%
As reviewed in Sec.~\ref{sec:4hspolIntroduction}, a number of studies
have attempted to locate this threshold value $\ratc$ of the polymer
size, leading to estimates of $\approx0.3\ldots 0.6$ in theoretical
work and $\approx0.25$ in experiments. The effect of colloid
polydispersity on $\ratc$ has remained unclear, however. To address
this point, we have investigated different pairs of values for
$(\rat,\pol)$ and determined the cloud curve for each of them. This is
computationally nontrivial because the equilibrium equations need to
be solved very accurately to determine the point at which a
fluid-solid binodal in a polydisperse system interchanges stability
with a gas-liquid binodal. Fig.~\ref{fig:xi_crossover} summarises our
results for two different colloid polydispersities.  For $\pol=0.01$,
we see that the gas-liquid binodal becomes unstable at
$\ratc=0.31\pm0.01$, while at $\pol=0.05$ this is shifted to the lower
value $\ratc=0.29\pm0.01$. We see that polydispersity {\em favours}
gas-liquid over fluid-solid phase separation, shifting the threshold
polymer size above which stable gas-liquid phase coexistence is
observed to lower values. Indeed, for a system with larger colloid
polydispersity ($\pol=0.1$, data not shown) we find a further
significant shift to $\ratc=0.25\pm0.01$. This is in good agreement
with the experimental data~\cite{IleOrrPooPus95} -- though for a
somewhat larger polydispersity than quoted in~\cite{IleOrrPooPus95} --
and is significantly below the value $\ratc=0.32$ found
in~\cite{LekPooPusStrWar92} for monodisperse colloids. In summary,
colloid polydispersity has significant effects on the threshold value
$\ratc$ for the polymer size, with a $\pol$ of only 10\% reducing it
from $\ratc \approx 0.31$ to $\ratc \approx 0.25$.

Several further comments are in order. First, for small $\pol$ one
would expect $\ratc$ to be decreased from its monodisperse value by
terms of order $\pol^2$. This is because phase boundaries should
generically shift by terms of this order~\cite{EvaFaiPoo98,Evans01} as
polydispersity is increased, while they should react smoothly, i.e.\
linearly, to changes in $\xi$. Unfortunately, we do not have at
present sufficient (and sufficiently precise) data to verify this
expectation. Second, our predicted $\ratc=0.31$ for $\pol=0.01$ is
just slightly smaller than the monodisperse value $\ratc=0.32$
from~\cite{LekPooPusStrWar92}. This could be due to the small
polydispersity in our case, though if shifts in $\ratc$ are indeed of
order $\pol^2$ then $\pol=0.01$ should give a shift 100 times smaller
than that ($0.31-0.25=0.06$) observed for $\pol=0.1$, which would be
negligible. More likely the small difference is due to the fact that
in~\cite{LekPooPusStrWar92} the scaled-particle or Percus-Yevick
expression was used for the free volume term $\muex\hs(\sigp)$ in
(\ref{eq:f_monopol}) while we employ the BMCSL expression,
consistently with the free energy $\fexhs$ describing the colloidal
fluid. Finally, we recall from Sec.~\ref{sec:cloud_shadow} that
colloid polydispersity {\em delays} the onset of gas-liquid
coexistence. On the other hand, we just saw that it {\em favours}
gas-liquid separation {\em relative} to fluid-solid. The conclusion is
that polydispersity must disfavour fluid-solid coexistence, and do so more
strongly than for gas-liquid. This is quite plausible, since
polydispersity causes inefficient particle packing in a crystalline
structure, while it can actually increase packing efficiency in a
liquid.

\subsection{Fractionation effects}
\label{sec:hspol_Fractionation_effects}

\begin{figure}
  \begin{center}
  \includegraphics[width=8cm]{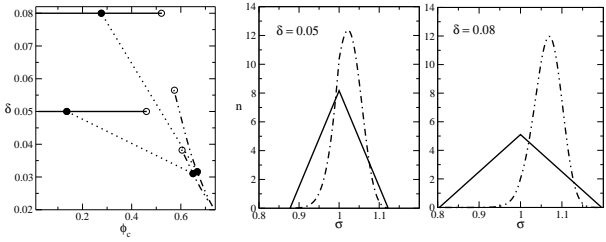}
  \caption{Left: Polydispersity $\pol$ of the fluid cloud and solid
  shadow phases versus their colloid volume fraction $\phic$, at the
  onset of F+S coexistence for $\rat=0.2$ and parent polydispersity
  $0.05$ and $0.08$. See Fig.~\ref{fig:d_xi_phip_vs_phic} (left) for
  the corresponding polymer volume fractions $\phip$. Solid lines
  refer to the fluid cloud phase, dot-dashed lines to the solid
  shadow; dotted lines connect sample cloud-shadow pairs. Middle and
  right: Example plots of the normalised colloid size distributions of
  the phases indicated by the circles in the left plot.
\label{fig:size_frac_flu_sol}}
  \end{center}
\end{figure}

As pointed out above, in colloid-polymer mixtures with polydisperse
colloids the onset of phase separation causes fractionation not just
between the colloids and the polymer, but also between the various
sizes of colloid particles. To demonstrate this,
Fig.~\ref{fig:size_frac_flu_sol} (left) shows how the polydispersities
of the fluid and solid phases vary along the cloud and shadow curves
of Fig.~\ref{fig:d_xi_phip_vs_phic} (left). On the $x$-axis we plot
the colloid volume fraction $\phic$ of the phases, so that the phases
are now represented in terms of $(\phic,\pol)$ instead of
$(\phic,\phip)$ as before. The corresponding values of $\phip$ can of
course be read off from the cloud and shadow curves in
Fig.~\ref{fig:d_xi_phip_vs_phic} (left). In particular, the ends of
the lines in Fig.~\ref{fig:size_frac_flu_sol} (left), marked by empty
circles, correspond to the limit $\phip\to 0$. Examples of the
normalised colloid size distributions, $n_{\rm c}(\sigc) =
\rhcsig/\int d\sigc\,\rhcsig$, are shown in
Fig.~\ref{fig:size_frac_flu_sol} (middle and right).
%
These results give a concrete demonstration of the fact that the
depletion interaction can be used to systematically control and reduce
the polydispersity of colloidal mixtures, as has been known for a
number of years. In particular Bibette~\cite{Bibette91} suggested a
procedure where colloid samples are fractionated from solutions of
volume fraction $\phic\approx0.1$. As we can see from
Fig.~\ref{fig:size_frac_flu_sol} (left), this is indeed the regime
where fractionation is strongest, i.e.\ where the shadow phase has a
much smaller polydispersity than the parent (which is identical
to the cloud phase). For phase separation at larger $\phic$,
corresponding from Fig.~\ref{fig:d_xi_phip_vs_phic} (left) to smaller
polymer volume fractions $\phip$, the reduction in polydispersity is
less pronounced.

Colloidal size fractionation effects do of course occur not only in
fluid-solid phase separation, but also in gas-liquid
coexistence. Evans {\it et al}~\cite{EvaFaiPoo98,Evans01} used a
perturbative approach to study such fractionation effects for
near-monodisperse parent size distributions. They predicted that the
difference in mean particle diameters of two coexisting phases,
$\Delta\bar{\sig}_{\rm c}=\bar{\sig}_{\rm c}^{(1)}-\bar{\sig}_{\rm
c}^{(2)}$, should universally scale as $\pol^\eta$ for small $\pol$,
with exponent $\eta=2$. In~\cite{FaiEva04} Fairhurst and Evans
verified this relation experimentally for a colloid-polymer mixture:
they used solutions of colloidal PMMA with random polystyrene coils,
with size ratio $\rat=0.45$. Collecting data from a number of samples
with parental colloid and polymer volume fractions in the region
$\phic=0.15\ldots 0.5$ and $\phip=0.15\ldots0.3$, they estimated a
power-law exponent $\eta=2.16 \pm 0.44$.

\begin{figure}
  \begin{center}
  \includegraphics[width=8cm]{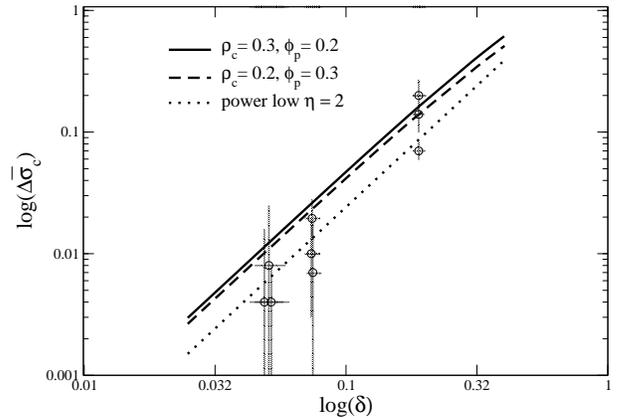}
  \caption{Log-log plot of the difference in mean colloid diameter
  between coexisting gas and liquid phases, as a function of the
  parent polydispersity. The polymer-colloid size ratio is
  $\rat=0.45$. Solid and dashed lines show our predictions for two
  different choices of colloid and polymer concentration, as indicated
  in the legend.  The dotted line is a power law with exponent
  $\eta=2$. The circles, together with grey error bars, indicate the
  experimental data points from~\cite{FaiEva04}, which were obtained
  from a set of samples in the range $\phic=0.15\ldots 0.5$,
  $\phip=0.15\ldots0.3$.
\label{fig:logDr_vs_logpol}}
  \end{center}
\end{figure}
%
To compare our numerical results with these experimental data, we
considered the same polymer-colloid size ratio $\rat=0.45$ and two
different choices for colloid and polymer concentrations,
$(\rhc=0.3,\phip=0.2)$ and $(\rhc=0.2,\phip=0.3)$, with colloid
polydispersity ranging from $\pol=0.025$ to $0.40$. We fix $\rhc$
rather that $\phic$ here since this is the case considered in the
perturbative theory. The colloid volume fraction $\phic=\langle
\sigc^3\rangle \rhc$ then increases with $\pol$ and lies between $0.2$
(for $\rhc=0.2$ and small $\pol$) and $0.45$ (for $\rhc=0.3$ and the
largest $\pol$). The overall range of variation of $\phic$ and $\phip$ is
thus comparable to but somewhat smaller than in the experiments
of~\cite{FaiEva04}. Fig.~\ref{fig:logDr_vs_logpol} shows our numerical
predictions for the difference in mean diameter of the coexisting gas
and liquid as a function of parent polydispersity, together with the
data from~\cite{FaiEva04}. As expected our calculations are consistent
with the universal quadratic scaling at small
polydispersities. More importantly, they also show
reasonable quantitative agreement with the experimentally measured
values. More precise experiments covering a narrower range of colloid
and polymer concentrations would obviously be useful, to permit a more
stringent test of our calculations.

\subsection{Inner phase boundaries\label{sec:hspol_Inner_phase_boundaries}} 

Having discussed the properties of the cloud and shadow curves, and
the fractionation effects which occur at or near the onset of phase
separation, we now turn to the internal phase boundaries and their
dependence on the polydispersity and polymer-colloid size ratio.

A general trend apparent from Fig.~\ref{fig:xi_d} is that, with
increasing polymer concentration, the phase boundaries shift
systematically to lower colloid volume fraction. Phase separation into
several solids, for example, can thus occur at much lower colloid
concentration than in a system without polymer. An intuitive
explanation for this is that the osmotic pressure in the system is
increased as polymers are added. This compresses the colloids and can
be viewed as effectively increasing the colloid volume fraction.  From
Fig.~\ref{fig:parent_hs}, phase equilibria involving several solids
will then occur earlier, as we observe.
Quantitatively, perhaps the most striking feature of
Fig.~\ref{fig:xi_d} (right) is the fact that for large polymers the
internal phase boundaries which continue those in the polymer-free
system are almost linear in the $(\phic,\phip)$-plane.  For the
boundaries of the G+L+S region in Fig.~\ref{fig:xi_d} (bottom right)
this is straightforward to rationalise. For monodisperse colloids, the
colloid-polymer mixture contains only two different particle species
and so the boundaries of the three-phase G+L+S triangle must be
{\em exactly} straight; compare Fig.~\ref{fig:pha_dia_sketch}. Introducing a
small degree of polydispersity $\pol$ should then leave these
boundaries approximately straight as observed. This explanation
already falters for Fig.~\ref{fig:xi_d} (top right), however, where
the G+L+S region widens rather than narrows as $\phip$ increases,
precluding a straightforward analogy with the three-phase triangle in
a monodisperse system. It fails completely for the inner phase
boundaries marking the transition between phases involving several
solids, which are induced purely by polydispersity effects and have no
monodisperse analogue. A different explanation is therefore required:
we will see that the near-linearity of the inner phase boundaries
arises from the fact that the gas phase consists almost entirely of
polymers, plus a negligible amount of colloids. It can therefore act
as a ``polymer buffer'' for the other phases, which contain only small
amounts of polymer and are otherwise similar to those in the pure
colloid system. For concreteness we focus below on a case where these
colloid-rich phases are solids, but the argument applies equally well
when there is a liquid among them.

\begin{figure}
  \begin{center}
  \includegraphics[width=8cm]{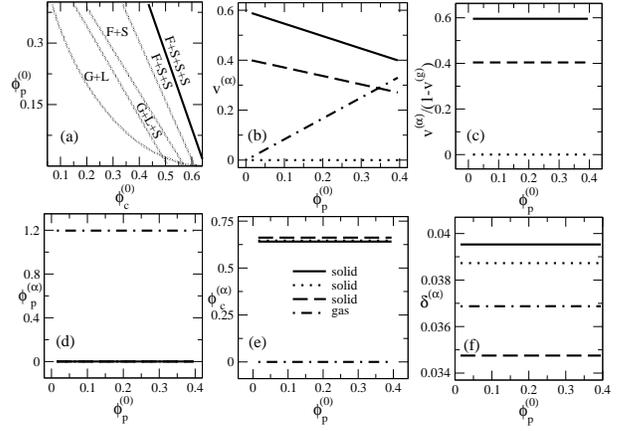}
  \caption{Properties of the coexisting phases, for a system with
  $\rat=0.4$ and $\pol=0.05$, along the phase boundary shown in (a) by
  the dark line. (b) Fractions of system volume $v^{(\alpha)}$
  occupied by the various phases; the newly forming solid has
  vanishing fractional volume. (c) Fractional volumes of the solids
  normalised by the total fractional volume occupied by solid
  phases. (d,e,f) Polymer and colloid volume fractions and colloid
  polydispersity of the coexisting phases.
\label{fig:properties}}
\end{center}
\end{figure}
To substantiate this hypothesis, we have extracted in
Fig.~\ref{fig:properties} the properties of the coexisting phases for
polymer size $\rat=0.4$ and polydispersity $\pol=0.05$, along the
phase boundary between the F+S+S and F+S+S+S regions highlighted by
the bold line in Fig.~\ref{fig:properties} (a). The position along the
boundary in the plots (b-f) is parameterised by the parent polymer
volume fraction $\phip^{(0)}$, i.e.\ the $y$-coordinate from (a). In
(d) and (e) we show the polymer and colloid volume fractions in the
various phases. We see that, as anticipated, the fluid phase consists
almost entirely of polymers, i.e.\ it is an extremely dilute colloidal
gas. The solids, on the other hand, are essentially pure colloid
phases and contain almost no polymer. In addition, as plots (e) and
(f) show, their properties remain unchanged along the phase boundary
and can be inferred from the extrapolation to the polymer-free limit
$\phip^{(0)}=0$. This applies in particular to their common pressure
$\Pi^*$, and the average colloid volume fraction in the solids,
$\phic^*$. The fractional system volumes $v^{(\alpha)}$ occupied by
the solids also need to remain in constant proportion to each other,
to maintain the overall colloid size distribution, consistent with the
results shown in Fig.~\ref{fig:properties}
(b). Figure~\ref{fig:properties} (c) shows, more explicitly, that when
we normalise the fractional volumes of the solid phases by the total
fractional volume of all solids, they become constant along the phase
boundary. The normalisation factor is $\sum_{\alpha\neq{\rm
g}}v^{(\alpha)} = 1-v^{(\rm g)}$, where ``g'' denotes the gas phase.

\begin{figure}
  \begin{center}
  \includegraphics[width=8cm]{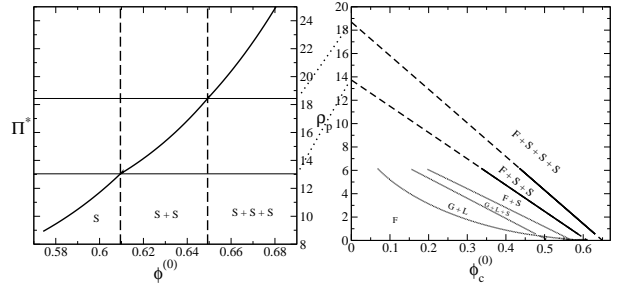}
  \caption{Left: Pressure plot of a polymer-free system with
  polydispersity $\pol=0.05$. The values of the pressure at the phase
  transitions are marked by the horizontal lines. Right: Phase diagram
  of a colloid-polymer mixture with the same colloid polydispersity
  and polymer size $\rat=0.4$, plotted as $\rhp^{(0)}$ vs
  $\phic^{(0)}$. Two phase boundaries are extrapolated (dashed) to
  $\phic^{(0)}=0$. The agreement between the extrapolated values of
  the intercepts, 13.7 and 18.7, and the equilibrium coexistence
  pressures $\Pi^*$ in the polymer-free system, 13.0 and 18.4, is
  good.
\label{fig:hspol_p_vs_phi_free-polymer}}
\end{center}
\end{figure}
We can now infer the shape of the phase boundary. Since only the solid
phases contain significant concentrations of colloid, particle number
conservation requires that the parental colloid volume fraction is given by
\begin{equation}
\phic^{(0)} = (1-v^{(\rm g)})\phic^*
\end{equation}
Similarly, because the polymers are found almost exclusively in the
gas, we have
\begin{equation}
\phip^{(0)} = v^{(\rm g)} \phip^{(\rm g)} = v^{(\rm g)}\rat^3\rhp^{(\rm g)}
\end{equation}
Here we have used the fact that in our density units
the polymer density $\rhp$ and volume fraction $\phip$ are related
simply by a factor $\rat^3$. In the same units, the pressure of the
gas phase is $\Pi^{(\rm g)} = \rhp^{(\rm g)}$ because the polymers are
ideal and the colloids make only a negligible contribution. At
equilibrium this pressure must equal that in the colloid phases,
$\Pi^*$, hence
\begin{equation}
\rhp^{(\rm g)} = \Pi^*
\end{equation}
Combining the last three equations now gives the desired relation
between the parent polymer and colloid volume fractions along the
phase boundary:
\begin{equation}
\rhp^{(0)} = \rat^{-3}\phip^{(0)} =
\Pi^*\left(1-\frac{\phic^{(0)}}{\phic^*}\right)
\label{eq:hspol_linear_g_dep}
\end{equation}
This is indeed linear, as we set out to show, and obeys the general
trend that with increasing $\phip^{(0)}$ phase transitions occur at
smaller parental colloid volume fractions $\phic^{(0)}$. The
relation~(\ref{eq:hspol_linear_g_dep}) also predicts that if we
extrapolate the straight phase boundaries to $\phic^{(0)} = 0$, i.e.\
to their intersection with the $y$-axis, the intercept will be
$\phip^{(0)}=\rat^3\Pi^*$, or $\rhp^{(0)}=\Pi^*$ if we plot
$\rhp^{(0)}$ on the $y$-axis. In
Fig.~\ref{fig:hspol_p_vs_phi_free-polymer} we check this prediction
explicitly for two of the numerically calculated phase boundaries of
the system in Fig.~\ref{fig:properties}, finding good agreement.
More generally, one concludes from~(\ref{eq:hspol_linear_g_dep}) that
all the straight phase boundaries in systems with sufficiently large
polymers should become approximately independent of the actual polymer
size $\rat$, once they are plotted in terms of $\rhp^{(0)}$ versus
$\phic^{(0)}$. The restriction here is that the polymers must not be
too close to the threshold size $\ratc$. This ensures that the
straight phase boundaries extend close to the polymer-free
baseline, thus justifying our extrapolation to $\phip=0$ to determine
$\Pi^*$ and $\phic^*$. For $\rat$
close to $\ratc$, on the other hand, as in e.g.\
Fig.~\ref{fig:high_pol_rho} above, the relevant polymer-free
``reference system'' can no longer be obtained by straightforward
linear extrapolation. Nevertheless, our arguments show that along
straight internal phase boundaries, and indeed along appropriate
straight lines {\em between} the phase boundaries, the colloidal size
distribution of the dense phases should remain approximately unchanged.

%
\begin{figure}
  \begin{center}
  \includegraphics[width=8cm]{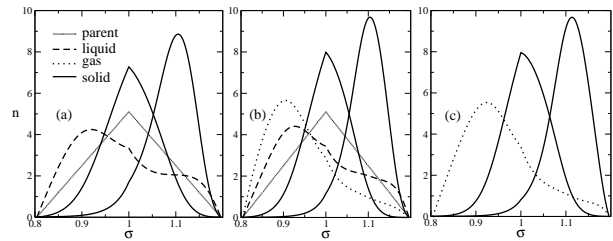}
  \caption{Normalised colloidal size distributions $n_{\rm c}(\sigc)$
  in coexisting phases; the parent distribution is shown for
  comparison. The three graphs correspond to the values of
  $(\phic,\phip)$ indicated in Fig.~\ref{fig:high_pol_rho}.
\label{fig:hspol_example_dis}
}
  \end{center}
\end{figure}
%
To illustrate this phenomenon, we plot in
Fig.~\ref{fig:hspol_example_dis} three examples of normalised colloid
diameter distributions at the three points marked in the phase diagram of
Fig.~\ref{fig:high_pol_rho}. These points lie approximately on a
straight line following the direction of the phase boundaries. As
expected, we find that the size distributions in the liquid and solid
phases remain approximately constant. This seems rather remarkable,
given that the overall colloid and polymer fractions in the three
cases are very different.

\section{\label{sec:4hspolConclusionandoutlook}Conclusion and outlook}

We have studied theoretically, for the first time, the equilibrium
phase behaviour of mixtures of {\em polydisperse} hard sphere colloids
and monodisperse polymers. Our treatment is based on the AO model,
which treats the polymers as spherical particles interacting only with
the colloids.
Within the van der Waals approximation of Lekkerkerker {\em et al},
the polymer-colloid interaction is determined by the excess chemical
potentials of the polydisperse hard sphere system. We have therefore
used as input suitable free energy approximations for the colloidal
fluid and solid, choosing the BMCSL and Bartlett's ``geometric'' free
energy, respectively; these were shown in previous work to give
quantitatively reliable predictions. Complete phase diagrams, taking
full account of fractionation effects, are found exactly -- to within
numerical accuracy -- by the use of the moment free energy with extra
adaptive moments; this is possible because the free energies involved
are truncatable. The intricate features of phase separation in pure
polydisperse colloids, including fractionation into several solids,
combine with the appearance of polymer-induced gas-liquid coexistence
to give a rich variety of phase diagram topologies.

We studied in some detail the dependence of these phase diagrams on
colloid size polydispersity $\pol$ and the polymer-colloid size ratio
$\rat$. Even for the moderate values of $\pol$ we consider ($\pol\leq
0.10$), polydispersity has a significant influence on the number of
phases and the topology of the phase boundaries. This influence arises
because the sequence of phase transitions that is observed in a
``baseline'' system of pure colloids changes significantly with
$\pol$. Starting from this baseline the addition of polymer then
causes fluid phases to phase separate into colloidal gas and liquid,
or a new fluid phase to appear if only solids were present in the
polymer-free baseline system. The strength of this effect grows with
polymer size $\rat$; this is reasonable because both the range and
strength of the effective colloid-colloid interaction caused by the
presence of polymer grows with $\rat$. The strength of this attraction
also grows with polymer concentration, and consequently all phase
boundaries shift to lower colloid concentrations when more polymer is
added. In particular, we predict that phase splits involving two or
more fractionated solids should occur at fairly moderate colloid
volume fractions, where the phase separation kinetics should be
substantially faster than for the corresponding higher volume
fractions in a system without added polymer. This suggests the
exciting possibility that such solid-solid phase splits, which have
not hitherto been seen in experiments, may be observable in this
regime. We note, however, that the growth kinetics of polydisperse
crystals~\cite{EvaHol01} may still cause deviations from the predicted
equilibrium behaviour even if the regime of multiple solids is
kinetically accessible.

We deduced the possible phase diagram topologies from the requirement
that they need to change smoothly with both $\pol$ and $\rat$. One
intriguing prediction of this is that, in the presence of colloid
polydispersity, stable three-phase gas-liquid-solid coexistence can
persist without a corresponding two-phase gas-liquid region in the
phase diagram; see the middle and top rows of
Fig.~\ref{fig:topologies}. This effect should also be observable
experimentally, for values of $\rat$ just below the threshold $\ratc$
at which the stable gas-liquid coexistence disappears.

In our quantitative analysis of the phase behaviour, we found that
this threshold value $\ratc$ depends significantly on colloid
polydispersity, decreasing from $\ratc\approx 0.31$ for $\pol=0.01$ to
$\ratc\approx 0.25$ for $\pol=0.1$. We saw that polydispersity delays
the onset of both gas-liquid and fluid-solid separation, but that the
effect is stronger for the latter, causing the observed shifts in
$\ratc$. We further considered fractionation effects and found that in
fluid-solid coexistence these are most marked for high polymer
and low colloid concentration, in agreement with experimental
protocols that are used 
to reduce polydispersity in colloidal systems. Also for gas-liquid
coexistence the calculated fractionation effects are in reasonable
agreement with the experimental data, and follow the universal
fractionation relation in the limit of small polydispersity. Finally,
we observed that, for sufficiently large polymer sizes $\rat$,
boundaries between regions of the phase diagram containing two or more
phases are close to straight. This is due to the occurrence of a dilute
colloidal gas phase which acts effectively as a polymer buffer.  We
derived from this an estimate of the location and slope of the phase
boundaries in terms of the properties of a corresponding polydisperse
colloid system without added polymer.

An interesting extension of this research would be to the case where
the {\em polymers} are polydisperse -- with a fixed parent density
distribution -- while the colloids are monodisperse, or even further
to the scenario where both colloids and polymers are
polydisperse. Work in this direction is in progress. One would also
like to assess the effects of polymer non-ideality, for which a number
of theoretical approaches have been developed, based on perturbation
theory around the $\theta$-point~\cite{WarIlePoo95}, integral
equations~\cite{AarTuiLek02} and effective colloid-colloid
interactions derived from simulations of self-avoiding walk (SAW)
polymers~\cite{LouBolMeiHan02}. The method that fits most naturally
into our framework is to add second-order virial terms to account for
polymer-polymer interactions, as done in~\cite{WarIlePoo95}. However,
recent simulation studies~\cite{BolLouHan02} indicate that the AO
model gives quantitatively reasonable results also for {\em
interacting} (monodisperse SAW) polymers up to quite large sizes
$\rat<0.34$. For larger $\rat$, where significant deviations
arise~\cite{BolLouHan02}, one has to account both for polymer
interactions {\em and} also increasingly for the detailed polymer
chain structure. An ``AO $+$ second virial'' model which only
addresses the first effect is therefore unlikely to be useful over a
significant range of $\rat$.

\addcontentsline{toc}{chapter}{Bibliography}
\bibliographystyle{aip}
\bibliography{bibliography_3hs,bibliography_lj,bibliography_hspol}

\end{document}